\documentclass[aps,prd,showpacs,nofootinbib,preprint]{revtex4}

\usepackage[latin1]{inputenc}

\usepackage{latexsym}
\usepackage{amsbsy}
\usepackage{mathrsfs}
\usepackage{color}

\usepackage{psfrag}

\usepackage{enumerate}

\usepackage{amsmath,amssymb,calc,amsfonts}
\usepackage{latexsym}
\usepackage{hyperref}

\usepackage{graphicx,calc,epsfig}
\def\ut#1{\rlap{\lower1ex\hbox{$\sim$}}#1{}}

\newcommand{\R}{\mathbb{R}}
\newcommand{\be}{\nopagebreak[3]\begin{equation}}
\newcommand{\ee}{\end{equation}}
\newcommand{\ba}{\nopagebreak[3]\begin{eqnarray}}
\newcommand{\ea}{\end{eqnarray}}
\DeclareFontFamily{U}{rsfs}{}         
\DeclareFontShape{U}{rsfs}{m}{n}{<5> rsfs5 <6><7> rsfs7          %
  <8><9><10><10.95><12><14.4><17.28><20.74><24.88> rsfs10}{}     %
\DeclareMathAlphabet{\mathfs}{U}{rsfs}{m}{n}                     %
\newcommand{\mfs}[1]{\mathfs {#1}}                               %

\newcommand{\va}{\scriptscriptstyle}

\newcommand{\sM}{{\mfs M}}

\newcommand{\sI}{{\mfs I}}\newcommand{\sO}{{\mfs O}}

\def\pb#1{\rlap{\lower1.5ex\hbox{$\longleftarrow$}}{#1}}
\def\dpb#1{\rlap{\lower1.5ex\hbox{$\Longleftarrow$}}{#1}}
\def\spb#1{\rlap{\lower1.5ex\hbox{$\leftarrow$}}{#1}}
\def\sdpb#1{\rlap{\lower1.5ex\hbox{$\Leftarrow$}}{#1}}

\definecolor{blue}{rgb}{0,0,1}
\definecolor{green}{rgb}{0,1,0}
\definecolor{red}{rgb}{1,0,0}
\definecolor{vio}{rgb}{1,0,1}
\definecolor{ama}{rgb}{1,1,0}

\begin{document}

%
%



\title{\bf Regular isolated black holes}

\date{\today}

\author{Carlos Kozameh$^1$, Osvaldo M. Moreschi$^1$ and Alejandro Perez$^2$} 

\affiliation{$^1$FaMAF, Instituto de Física Enrique Gaviola (IFEG), CONICET, \\
Ciudad Universitaria, (5000) C\'ordoba, Argentina. }

\affiliation{$^2$Centre de Physique Th\'eorique,
Campus de Luminy, 13288
Marseille, France.}

\begin{abstract}

We review a recently introduced\cite{Kozameh10} approach to study spacetimes which contain an
isolated black hole. We allow for gravitational radiation to fall through
the event horizon and/or to be radiated at infinity.
The crucial idea in our setting is the introduction of physical coordinates
that have the meaning of being null hypersurfaces near the horizon and
in a neighborhood of future null infinity.
These spacetimes have remarkable properties, as for example 
a generalized notion of  surface gravity $k_{H}$.

\end{abstract}

\maketitle

\section{Introduction}

Stationary black holes have been thoroughly studied in the literature and exhibit important
geometrical features; as is the notion of surface gravity.
However, since astrophysical black holes may absorb matter and/or radiation, and
also may emit gravitational radiation, it is important to know whether the
basic properties of stationary black holes are shared  by non-stationary ones.

In this work we treat this problem recurring to a geometrical setup;
that allows us to study the
late stage of dynamical evolution once all the matter has fallen into the black hole but still 
taking into account incoming and outgoing radiation.

{ Our analysis is based on the assumption that the null surfaces associated to suitably chosen (inertial) retarded time $u$ at future null infinity
define a smooth null foliation in the vicinity of the black hole event horizon.
More precisely,  there is a null physical coordinate $w(u)$ (unique up to scaling) that can be used to describe fields in the vicinity of the horizon. 
The above assumption, made precise in what follows, characterizes  the spacetimes studied in this letter which are referred to as  {\em isolated black holes} (IBH). 

In this work we obtain an
exponential relationship between the null function $w$ and a preferred retarded time $u$ function. 
We also show that
there exists a geometrically defined smooth vector field that is null both at the horizon $H$ and at future null infinity.
There is also a generalized notion of surface gravity  which is constant on $H$.

}

\begin{figure}[h]
\includegraphics[clip,width=0.48\textwidth]{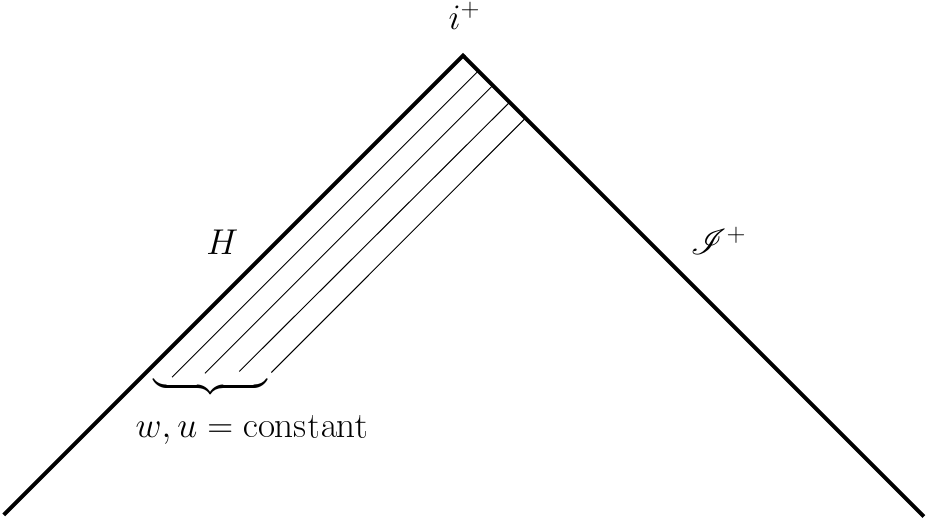}
\caption{An isolated black hole spacetime. {\sf  In a vicinity of $i^+$, the null surfaces of constant retarded time $u$ 
are smooth all the way up to the event horizon for IBHs.}}
\label{figui}
\end{figure}

\section{The geometrical setting}

Consider $(\sM,g_{ab})$ an asymptotically flat spacetime at future null infinity containing a black hole.
Its conformal diagram is depicted in Figure \ref{figui}. In  the past of an open set of future 
null infinity ($\sI^+$)---defined by those points for which
their Bondi
%
 retarded time $u$ 
 is in the range $u \in (u_0, \infty)$---we require the existence of a regular null function $w(u)$ 
such that: $w=0$ at the horizon $H$,  and $w<0$ in the region of interest.
Clearly there is a large freedom in selecting the function $w(u)$ if the only requirement is to satisfy the
above condition. 
In what follows we completely reduce this freedom by selecting a unique (up to constant scaling) null function $w(u)$ 
using the available structure at $\sI^{+}$.  We can now list the ingredients of our construction:

$(\sM,g_{ab})$ is an asymptotically flat spacetime at future null infinity containing a black hole,
assumed to be vacuum in the vicinity of $i^{\va +}$.
The topology of the BH event horizon  
$H$ is assumed to be $S^2\times \R$ in that region.

We assume the existence of a null function $u$ in the asymptotic region such that
the cuts at $\sI^{\va +}$ are Bondi. We introduce the one form $\tilde \ell_a\equiv (du)_a$, then the vector field $\tilde \ell^a$ defines a
null geodesic congruence.  It is therefore
natural to introduce the affine function $r$ through
$\tilde \ell=\frac{\partial}{\partial r}$.

In a space-time satisfying the previous conditions there is a three parameter family of Bondi systems such that the leading order behaviour of the shear 
of the congruence defined by $\tilde \ell$ vanishes in the limit $u\to \infty$.  We completely eliminate this freedom by
requiring that the cuts $u=$constant asymptotically coincide
with the {\em center of mass}
in the same regime.%

The functions $(u,r)$ can be used as coordinate functions in the region
in which the null congruence $\tilde \ell$ does not show caustics.
The asymptotically flat region is reached in the limit
$r \rightarrow \infty$. The origin of this affine parameter is chosen so that $r$ asymptotically 
coincides with the luminosity distance (i.e. the expansion of the null congruence $\tilde \ell$ 
goes as $r^{-1}+\sO(r^{-3})$).

There exists a smooth null function $w=w(u)$ such that
\begin{enumerate}
\item $w=0$ at the horizon $H$.

\item $\dot w \equiv \frac{dw}{du}>0$.

\item $w < 0$ for all $u$.

\item $\lim\limits_{u\rightarrow \infty} w = 0$.
\end{enumerate}

We define the one form $\ell_a\equiv (d w)_a$, then,
the geodesic vector field $\ell^a$  is also tangent to the 
null congruence defined in point 2. It is therefore
natural to introduce the affine function $y$ through
$\ell=\frac{\partial}{\partial y}$.

The functions $(w,y)$ can be used as coordinate functions in the region
where  the null congruence $\ell$ does no show caustics.

\label{veamos}Given the functions $(u,r)$ one can choose the affine
parameter $y$, for each null geodesic,
so that the 2-surfaces $u$=const., $r$=const. coincide with the
 2-surfaces  $w$=const., $y$=const. This implies the following relationship between $r$ and $y$:
\be
r=\dot w y + r_{0}(w)
\label{eqq}.
\ee
One then completes a coordinate system by choosing angular sphere-coordinates $(\theta,\phi)$
or stereographic sphere-coordinates $(\zeta,\bar \zeta)$.

In the exterior of the BH there is clearly a smooth relationship between the pairs $(u,r)$ and $(w,y)$. We assume that $r$ is a smooth function of $(w,y)$ 
all the way up to the horizon.


Let us observe that
from the null vector fields $\ell^a$ and $ \tilde \ell^a$ one can construct null tetrads
$(\ell^a,m^a, \bar m^a, n^a)$, and $(\tilde \ell^a, \tilde m^a, \bar\tilde m^a,\tilde n^a)$ adapted to the
geometry of the coordinate system introduced above.
The freedom in this choice is reduced by choosing  the vectors $m^a= \tilde m^a$ and  tangent to
the topological 2-spheres $(w,y)=\text{constant}=(u,r)$.

From $w=w(u)$ it follows that $dw=\dot w \,du$ which implies the following relation between the two tetrads
\be\ell^a=\dot w\  \tilde \ell^a, \ \ \  n^a=\frac{ 1}{\dot w}\tilde n^a, \ \ \ m^a= \tilde m^a. \label{mainy} \ee
If we denote the five (complex) Weyl tensor null tetrad components\cite{Geroch73}  $\Psi_{N}^{}$ and $\tilde \Psi_{N}^{}$ for $N\in \{0,1,2,3,4\}$
in each of the respective tetrads, then we get the following relations
\begin{equation}\label{eq:Psi-N}
 \Psi^{}_N = \dot w^{(2-N)}  \tilde \Psi^{}_N.
\end{equation}

Before proceeding further we would like to give a motivation for some of the above
assumptions.
From the previous relation one has
\begin{equation}
 \Psi^{}_2(w,y,x^A)=\tilde \Psi_2^{}(u(w),r(w,y),x^A);
\end{equation}
where $(x^A)$ denotes angular coordinates.
Therefore, since the left hand side is a regular expression in terms of the coordinate $w$,
it is natural for $r$ to be a regular function of $w$ as it was assumed   above.

\section{Main results}

The first consequence of the regularity property is that the limit
 $r_H \equiv \lim_{w\rightarrow 0} r(w,y)$ exists.
Another consequence  is that
 $\dot w(w)$ and  $r_0(w)$, as defined in equation (\ref{eqq}), are regular functions of $w$. Moreover, we have 
\be
w(u)=\int\limits^{u}_{\infty} \dot w(u') du' \ \ \
\Rightarrow\ \ \ \dot w(0)=\lim\limits_{u\to \infty} \dot
w(u)=0.\label{limit}
\ee 
 It then follows from (\ref{eqq}) that
\begin{equation}
 r_H = \text{constant} .
\end{equation}
This allows to define $r_H$ as the radius of the isolated black
hole. Since by assumption $\dot w(w)$ admits a Taylor expansion around $w=0$ we can write. 
\be 
\dot w =a w+ \sO(w^2).
\ee 
 Assuming that $a\not=0$ the above equation can be integrated giving the
important relation 
\be\label{main} 
\boxed{
w(u)=-\exp{(a(u-u_0))}+\sO(\exp{(2au)}) }  ,
\ee
where $\exp(-a u_0)$ is the rescaling freedom mentioned previously associated with the choice of origin for
the Bondi retarded time $u$.
Note also that in order to satisfy (\ref{limit}) one has $a<0$.

A natural question arises: is it possible to get
more information concerning the nature of the coefficient $a$? Next we show that $a$ has
a clear geometrical meaning.
We start from  the vector field
\begin{equation}
\chi \equiv \frac{\partial}{\partial u}.
\end{equation}
The vector field $\chi$ has several useful properties:
It is a smooth vector field that is a null geodesic generator at $\sI^{+}$. As $u$ is a Bondi coordinate 
it generates inertial time translations at future null infinity.


It is a null geodesic generator of the horizon $H$.

At the horizon $H$, $\chi$ satisfies the equation,
\begin{equation}\nonumber
\chi^{a}\nabla_{a}\chi^{b} \equiv k_H \chi^{b} ;
\end{equation}
where $k_H$ is a  generalized surface gravity.

\ \ \ \ \ \ \ \ \ \ \ \ \ \ \ \ \ 
\be\boxed{ k_H = \text{const.} =-a }.\ee


The  previous statements follow from expressing $\chi$ in terms of the regular
coordinates $(w,y,\zeta,\bar{\zeta})$, namely
\begin{align}\nonumber 
\chi  & =\dot{w}\frac{\partial}{\partial w}%
-\frac{\partial r}{\partial w}\frac{\partial}{\partial y}\\
& =aw\frac{\partial}{\partial w}-(ay+\frac{dr_{0}}{dw})\frac{\partial
}{\partial y}+{\mathfs {O}}(w^{2}). \label{cucu}
\end{align}
Evaluating  the previous equation at $w=0$ one obtains
\begin{equation}\label{pito}
\left.  \chi\right\vert _{w=0}=-\left(ay+\left.\frac{dr_{0}}{dw}\right|_{\va w= 0}\right)\frac{\partial
}{\partial y}
,
\end{equation}
Eq. (\ref{pito})  implies properties 2 to 4.

In this way, the class of spacetimes considered here admits  a notion of surface gravity which 
coincides with the usual one in cases when the
spacetime is stationary, e.g. a member of the Kerr family. Note that
if we had taken $a=0$ above, one would have obtained $k_{H}=0$. This case corresponds to the especial cases 
involving (in particular) the stationary
extremal black holes.

With this definition of surface gravity, the relation between
the null coordinate $w$ and the Bondi retarded time $u$ reads%
\begin{equation}\label{eq:expdecay}
\boxed{
w=-\exp{(-k_{H}(u-u_{0}))}+{\mathfs {O}}(\exp{(-2k_{H}u)})
};
\end{equation}
which we recognize as the generalization of the smooth Kruskal coordinate
transformation that can be found in Schwarzschild and Kerr geometries.

Finally, we want to address the issue of characteristic data in our framework;
which is $\Psi_0(y,\theta,\phi)$ at $H$ and $\Psi_4^0(w,\theta,\phi)$
at $\mathscr{I}^+$.
At the horizon $H$ one has to choose $\Psi_0$ going to zero as
$y\rightarrow \infty$.
A necessary condition for this behaviour is that $\Psi_0$ goes to zero at least as $O(1/y^3)$.
This comes from the requirement that the area of the sections
$y$=constant must
go to a finite value when $y\to \infty$.
At future null infinity a physical condition is that the radiation
field $\tilde\Psi_4^0$
must go to zero as  $u\rightarrow\infty$. 

However, our characteristic  data $\Psi_4^0$ could be unbounded as $w\to 0$;
since it can be shown that it is related to the radiation field  via $\Psi_4^0 =\tilde\Psi_4^0/ \dot w^3$. 
This is the case for radiation fields 
predicted in studies of gravitational collapse in perturbation theory where $\tilde\Psi_4^0$ has a power-law fall-off 
behaviour as $u\to \infty$ \cite{Gundlach94}.
It might appear that such unbounded data could invalidate some of the assumptions in the definition of an IBH.
Nevertheless, just from the physical condition that $\tilde\Psi_4^0$ vanishes
as  $u\rightarrow\infty$, one can prove that
%
the null geodesic congruence used in our construction is caustic free in the region of interest. Therefore, the power law
fall-off behaviour for the characteristic data expected in gravitational collapse
is admitted in our framework.

\section{Properties of the null coordinate system}
The basic setting assumes the existence of regular null hypersurfaces near the horizon,
that can reach out up to future null infinity.
Coming from the asymptotic region to the interior, in a general IBH, one would expect
at some point to encounter a caustic of the null geodesic congruence previously defined.
Our constructions is good up from the asymptotic region
up to the appearance of such caustics.
We therefore study here, the condition that these assumptions impose on the spacetimes.

For this purpose we study the appearance of caustics from the study of the
behavior of the optical scalars,
which, in vacuum, satisfy the equations
\begin{equation}\label{eq:thornrho-l}
\frac{\partial \rho}{\partial y}
=
\rho ^{2}
+\sigma \, \bar\sigma
,
\end{equation}
\begin{equation}\label{eq:thornsigma-l}
\frac{\partial \sigma}{\partial y}
=
2 \rho \, \sigma
+\Psi_0
.
\end{equation}

Assuming the boundary conditions
$\rho(y=\infty)=0$ and $\sigma(y=\infty)=0$,
one can be expressed the above equations in integral form, by:
\begin{equation}\label{eq:rhoint}
\rho(y)
=
\int_{\infty}^y
\left(
\rho ^{2}
+\sigma \, \bar\sigma
\right) dy
 ,
\end{equation}
\begin{equation}\label{eq:sigmaint}
\sigma(y)
=
\int_{\infty}^y
\left(
2 \rho \, \sigma
+\Psi_0
\right)
dy
.
\end{equation}

These equations can be solved by means of the recursive relation
\begin{equation}\label{eq:rhoimas1}
\rho_{i+1}(y)
=
\int_{\infty}^y
\left(
\rho_i^{2}
+\sigma_i \, \bar\sigma_i
\right) dy
 ,
\end{equation}
\begin{equation}\label{eq:sigmaimas1}
\sigma_{i+1}(y)
=
\int_{\infty}^y
\left(
2 \rho_i \, \sigma_i
+\Psi_0
\right)
dy
.
\end{equation}
For the first step in the iteration we take the solution $\rho^*$ and $\sigma^*$
of the system for $\Psi_0=0$; in other words we take
$\rho_{i=0} = \rho^*$ and $\sigma_{i=0} = \sigma^*$.

The solution of the above equations when $\Psi_0=0$  are
\begin{equation}
\rho^* = -\frac{y+y_0(w)}{(y+y_0(w))^2-\sigma_0(w) \bar\sigma_0(w)},
\end{equation}
\begin{equation}
\sigma^* = \frac{\sigma_0(w)}{(y+y_0(w))^2-\sigma_0(w) \bar\sigma_0(w)}.
\end{equation}

It is probably worthwhile to recall that
\begin{equation}
y_0 = \frac{r_0}{\dot w},
\end{equation}
(since we are using $r = \dot w(y+y_0)$) and
\begin{equation}
\sigma_0 = \frac{\tilde\sigma_0}{\dot w}
;
\end{equation}
where $\tilde\sigma_0$ determines the Bondi radiation data.

Also, let us observe that for finite $r_0(0)$ and standard behavior of $\tilde\sigma_0$,
one has, for any value of $y$, that $\lim_{w \to 0} \rho^* = 0$ and $\lim_{w \to 0} \sigma^* = 0$.

It can be seen then that the Bondi radiation data at future null infinity;
which can be encoded in $\tilde\Psi_4^0$(since $\tilde\Psi_4^0=-\ddot{\tilde\sigma}_0$), 
and the knowledge of $\Psi_0$
determine completely the optical scalars.

It is clear that this iterative procedure will give the solution if it converges.
This obviously impose conditions on the asymptotic properties of $\Psi_0$ (for $y\to\infty$)
and $\tilde\Psi_4^0$ (for $u\to\infty$).
We have seen previously that at the horizon one needs $\Psi_0$ to behave as $O(1/y^3)$;
so that just from the continuity of $\Psi_0$ in a neighborhood of the horizon $H$
one concludes that this iterative procedure provides solutions where the location
of the caustics will determine continuous sets.


%

Therefore since one expects caustics at the horizon, let say at values
$y_{Hc}$, one also expects caustics near the horizon at values $y_c$ close to $y_{Hc}$.

Although we do not calculate in this work the explicit asymptotic behavior for the two characteristic
data that allow our construction, it is clear that it is just a matter of finding the
appropriate fall off of these fields.
In future works we will study whether our formulation allows for the expected
power law fall off behavior characteristic of tails; found in perturbative studies
of stationary black holes.

\section*{Summary and final comments}

In this work we have introduced a framework to study black holes in their late 
stage of dynamical evolution, that is, after all matter has fallen into the BH but 
still taking into account both incoming and outgoing radiation. Our framework is sufficiently 
general to include physically interesting collapsing scenarios.
The construction of two related  geometrical null coordinate systems adapted to the available structure, 
plus mild regularity conditions at the horizon lead to novel results: physical coordinates (and associated 
null tetrads) to study the field equations in the late phase of collapse,  a generalized  surface gravity which is constant (zeroth law of black hole mechanics), the correct exponential behaviour between $w$ and $u$ (mediated by the surface gravity)
which should be relevant for the study of Hawking BH radiation.

In future  work we will explicitly analyze the Einstein field equations near the horizon $H$,
expecting to obtain more dynamical information describing the late phase of gravitational collapse.

\subsubsection*{Acknowledgements}
We acknowledge financial support from CONICET, SeCyT-UNC, Foncyt and by the Agence Nationale de la Recherche; grant
ANR-06-BLAN-0050. A.P. was supported by {\em l'Institut Universitaire de France}.


\providecommand{\href}[2]{#2}\begingroup\raggedright\endgroup

\end{document}